\pdfoutput = 1
\documentclass[aps,prl,twocolumn,showpacs,superscriptaddress,10pt]{revtex4-1}

\usepackage{graphicx}
\usepackage{color}
\usepackage{amsmath}
\usepackage{amsfonts}
\usepackage{bm}
\usepackage{enumerate}
\usepackage{array}

\newcommand{\abs}[1]{\ensuremath{\lvert #1 \rvert}}

\newcommand{\avg}[1]{\ensuremath{\langle #1\rangle}}

\begin{document}

\title{Majorana-Like Modes of Light in a One-Dimensional Array of Nonlinear Cavities}

\author{C.-E. Bardyn}
\affiliation{Institute for Quantum Electronics, ETH Zurich, 8093 Zurich, Switzerland}

\author{A. \.Imamo\u glu}
\affiliation{Institute for Quantum Electronics, ETH Zurich, 8093 Zurich, Switzerland}

\begin{abstract}

The search for Majorana fermions in $p$-wave paired fermionic systems has recently moved to the forefront of condensed-matter research. Here we propose an alternative route and show theoretically that Majorana-like modes can be realized and probed in a driven-dissipative system of strongly correlated photons consisting of a chain of tunnel-coupled cavities, where $p$-wave pairing effectively arises from the interplay between strong on-site interactions and two-photon parametric driving. The nonlocal nature of these exotic modes could be demonstrated through cross-correlation measurements carried out at the ends of the chain---revealing a strong photon bunching signature---and their non-Abelian properties could be simulated through tunnel-braid operations.

\end{abstract}


\maketitle



In recent years, strongly correlated photons have proved to be a remarkably rich platform for investigating phenomena traditionally regarded as pertaining to condensed matter physics. Tremendous theoretical and experimental efforts have made it possible to achieve strong optical nonlinearities at the single-photon level~\cite{Brune96, Fink08} and to demonstrate photon blockade effects~\cite{Imamoglu97, Birnbaum05, Faraon08, Bozyigit11, Hoffman11, Reinhard12}. Meanwhile, the pursuit of Majorana fermions has become a new focus of condensed matter research~\cite{Reich12, Alicea12}, and $p$-wave paired superfluids and superconductors have been promoted as paradigmatic systems for investigating the physics of Majorana modes~\cite{Read00}. Even though strongly interacting photons have been predicted to exhibit a typical fermionic behavior~\cite{Chang08, Carusotto09}, optical systems have never been considered as candidates for realizing such exotic physics.

In this Letter, we show that Majorana-like modes (MLMs) can be obtained in a one-dimensional (1D) strongly correlated system of impenetrable (or ``fermionized'') photons. More specifically, we consider a chain of coupled cavities with strong on-site nonlinearities and introduce a drive mechanism based on parametric amplification which, in stark contrast to previous works, gives rise to an effective $p$-wave pairing between (fermionized) photons. We map our system to the 1D chain originally proposed by Kitaev as a toy model for Majorana fermions~\cite{Kitaev01}, and show the existence of zero-energy modes with properties similar to those of Majorana modes in solid-state systems. Owing to the intrinsic dissipative nature of the system, these ``Majorana-like'' modes do not benefit from parity (or ``topological'') protection against decoherence~\cite{Trauzettel11, Bardyn12}, and thus cannot serve as topological quantum memories~\cite{Kitaev03}. However, they do behave as genuine Majorana modes on time scales shorter than the lifetime of a photon in the system, allowing Majorana physics to be probed.

To demonstrate the fact that MLMs can be detected via simple optical schemes, we propose a realistic experiment that takes full advantage of the optical nature of the system and allows for the direct observation of MLMs through second-order photon cross-correlation measurements. Although our proposal is strictly limited to 1D---since impenetrable photons do not behave as fermions in higher dimensions---we show that MLMs can effectively be exchanged using ``tunnel-braid'' operations~\cite{Flensberg11}, enabling us to simulate their non-Abelian properties.

\begin{figure}[b]
    \begin{center}
        \includegraphics[width=\columnwidth]{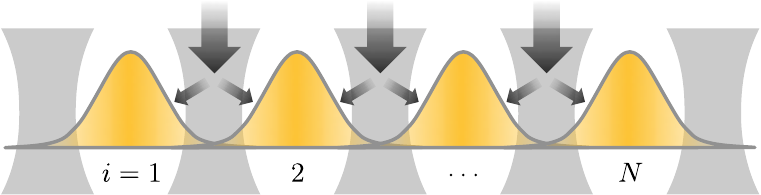}
        \caption{(Color online). Driven-dissipative chain of coupled cavities. Each cavity exhibits a large optical nonlinearity and sustains a single Wannier mode which weakly overlaps with its nearest neighbors, thus allowing for photon hopping between sites. Parametric pumps (depicted by arrows) couple to the weak inter-cavity field and inject photon pairs which, owing to strong photon-photon repulsion, split up into different cavities, effectively giving rise to $p$-wave pairing.}
        \label{fig:system}
    \end{center}
\end{figure}


\emph{The model.}---The backbone of our system consists of a 1D chain of $N$ optical cavities coupled through nearest-neighbor photon tunneling [Fig.~\ref{fig:system}]. Each cavity exhibits a large optical nonlinearity (i.e., is strongly coupled to an artificial atom) and supports a single mode that can be described as a Wannier function localized on site $i$ around the cavity center. Photon tunneling occurs as a result of the non-vanishing spatial overlap between nearest-neighboring Wannier modes~\cite{Hartmann06}, and the system Hamiltonian takes the generic form of a generalized Bose-Hubbard model:
\begin{align} \label{eqn:boseHubbard}
    H_0 & = \omega_c \sum_{i=1}^{N} b^\dagger_i b_i + \tfrac{U}{2} \sum_{i=1}^{N} b^\dagger_i b^\dagger_i b_i b_i - J \sum_{i=1}^{N-1} (b^\dagger_i b_{i+1} + h.c.),
\end{align}
where $b_i$ ($b^\dagger_i$) are annihilation (creation) operators associated with the $i^{\text{th}}$ cavity of the chain with resonance frequency $\omega_c$, $U$ is the strength of the on-site photon-photon repulsion (Kerr energy) due to the large optical nonlinearities, and $J$ denotes the tunneling amplitude between nearest-neighboring sites. In this work, we will focus exclusively on the strong-interaction regime, in which the energy cost $U$ of adding an extra photon to an occupied cavity is by far the largest of all relevant energy scales~\cite{Footnote:rotatingFrame}. In this so-called ``hard-core'' limit, the occupation of each site is effectively restricted to $0$ or $1$, and the photons exhibit a characteristic fermionic behavior~\cite{Chang08, Carusotto09}.

To achieve $p$-wave pairing, we introduce \emph{parametric} pumps (or amplifiers) which, in stark contrast to usual coherent drives, are tailored to inject \emph{pairs} of photons into the system through nonlinear optical processes (see Supplemental Material~\cite{SupplementalMaterial}). Assuming that these pumps drive the system locally through the inter-cavity field---which consists of a superposition of two neighboring Wannier modes---photons from a single pair can either be emitted into the same cavity (or Wannier mode), or settle into different, nearest-neighboring cavities. In the strong-interaction regime, the second process is strongly favored, and the drive Hamiltonian effectively reads
\begin{align} \label{eqn:driveHam}
    H_{\text{drive}} & = - \abs{\Delta} \sum_{i=1}^{N-1} (e^{\mathrm{i} (2 \omega_p t + \phi)} b_i b_{i+1} + h.c.),
\end{align}
where $\Delta = \abs{\Delta} e^{\mathrm{i} \phi}$ defines the amplitude and phase of the parametric pumps, and $\omega_p$ their frequency. We note that the amplitude $\abs{\Delta}$ of the parametric drive is determined by the overlap of the Wannier modes in a similar way as the tunneling amplitude $J$ defined above. We thus expect to be able to reach a regime in which the two quantities are of the same order. Physically, the above Hamiltonian describes the coherent exchange of \emph{$p$-wave} paired photons between the system and the classical pump field(s)~\cite{Footnote:pWavePairing}. It provides the optical counterpart of $p$-wave superconductivity that is crucial to access Majorana physics, and compensates for losses by continuously replenishing the system with photons. The time evolution of the system including the drive and the photon losses is governed by the Lindblad master equation
\begin{align} \label{eqn:masterEquation}
    \partial_t \rho & = - \mathrm{i} \left[ H_0 + H_{\text{drive}}, \rho \right] + \Gamma \sum_{i=1}^N \left( b_i \rho \, b_i^\dagger - \tfrac{1}{2} \{ b^\dagger_i  b_i, \rho \} \right),
\end{align}
where $\rho$ is the density matrix of the system and $\Gamma$ the decay rate associated with the individual cavities.


\emph{Mapping to a 1D Kitaev chain.}---In the strong-interaction regime ($U \gg J, \abs{\Delta}$), the Hilbert space of the system effectively reduces to that of hard-core photons $\tilde{b}_i = \mathcal{P} b_i \mathcal{P}$, $\tilde{b}^\dagger_i = \mathcal{P} b^\dagger_i \mathcal{P}$, where $\mathcal{P}$ projects onto the subspace of single occupancy. Hard-core photons can be seen as spin-$1/2$ particles, with Pauli-type matrices $\sigma^{-}_i = 2 \tilde{b}_i$, $\sigma^{+}_i = 2 \tilde{b}^\dagger_i$ ($\sigma^{\pm}_i = \sigma^{x}_i \pm \mathrm{i} \sigma^{y}_i$), and their fermionic nature can be unveiled by mapping the spin-$1/2$ particles to spinless fermions $a_i$, $a^\dagger_i$ using a Jordan-Wigner transformation~\cite{Cazalilla11} of the form $a_i = \tfrac{1}{2} \prod_{j=1}^{i-1} (-\sigma^{z}_j) \sigma^{-}_i$. Defining $\mu = \omega_p -\omega_c$ and moving to a rotating frame defined by $H_1 = \omega_p \sum_i \tilde{b}^\dagger_i \tilde{b}_i$, the Hamiltonian $H = H_0 + H_{\text{drive}}$ of the full system becomes, in the fermionic picture,
\begin{align} \label{eqn:kitaevHam}
    H = & - J \sum_{i=1}^{N-1} (a^\dagger_i a_{i+1} + h.c.) + \abs{\Delta} \sum_{i=1}^{N-1} (e^{\mathrm{i} \phi} a_i a_{i+1} + h.c.) \nonumber \\
    & - \mu \sum_{i=1}^{N} a^\dagger_i a_i,
\end{align}
which corresponds to the 1D $p$-wave superconductor of spinless fermions originally introduced by Kitaev~\cite{Kitaev01}. Here, the cavity frequency $\omega_c$ plays the role of a Fermi level, and the detuning $\mu = \omega_p -\omega_c$ between the pump and cavity frequencies that of a chemical potential. Assuming that $\abs{\Delta} \neq 0$, two topologically distinct (gapped) phases can be identified~\cite{Kitaev01}: a trivial phase corresponding to $\abs{\mu} > \abs{2J}$, and a nontrivial phase corresponding to $\abs{\mu} < \abs{2J}$, in which the system supports Majorana modes that are exponentially localized at both ends of the chain. The topological phenomena associated with the 1D Kitaev chain can be most easily understood for $J = \Delta > 0$ and $\mu = 0$ (i.e. $\omega_p = \omega_c$). In this illustrative case, the Hamiltonian reduces to
\begin{align} \label{eqn:idealKitaevHam}
    H = \mathrm{i} J \sum_{i=1}^{N-1} c_{2i} c_{2i+1} = - J \sum_{i=1}^{N-1} \sigma^{x}_i \sigma^{x}_{i+1},
\end{align}
with Majorana operators defined as
\begin{align}
    \setlength{\extrarowheight}{3pt}
    \begin{array}{lll}
        c_{2i-1} & = \phantom{-\mathrm{i}(} a_i + a^\dagger_i & = \phantom{-} \prod_{j=1}^{i-1} (-\sigma^{z}_j) \sigma^{x}_i, \\
        c_{2i} & = -\mathrm{i} (a_i - a^\dagger_i) & = -\prod_{j=1}^{i-1} (-\sigma^{z}_j) \sigma^{y}_i,
    \end{array}
\end{align}
and can readily be diagonalized as $H = 2 J \sum_{i=1}^{N-1} (\tilde{a}^\dagger_i \tilde{a}_i - 1/2)$ with Bogoliubov-Valatin quasiparticle operators $\tilde{a}_i = (c_{2i} + \mathrm{i} c_{2i+1})/2$. The associated spectrum is symmetric about the ``Fermi level'' $\omega_c$ and features a gap $2J$. Most importantly, it exhibits two Majorana zero-energy modes corresponding to the Majorana operators $c_1$ and $c_{2N}$ localized at the ends of the chain but absent from the Hamiltonian. These modes define a two-dimensional, \emph{nonlocal} degenerate (zero-energy) subspace which we identify as a ``Majorana qubit'', with associated $\sigma^{z}$ operator
\begin{align} \label{eqn:sigmazM}
    \sigma^{z}_M = \mathrm{i} c_1 c_{2N} = \prod_{j=1}^{N} (-\sigma^{z}_j) \sigma^{x}_1 \sigma^{x}_N.
\end{align}
The string-like operator $P = \prod_{j=1}^{N} (-\sigma^{z}_j)$ that appears in the above expression corresponds to the parity operator associated with the \emph{total} number of (fermionized) photons. It commutes with the Hamiltonian of Eq.~\eqref{eqn:kitaevHam}, but \emph{anticommutes} with the collapse operators $\sigma^{-}_i$ entering the Liouvillian in Eq.~\eqref{eqn:masterEquation}. Physically, this means that single-photon losses result in the breakdown of parity conservation, as expected, such that the Majorana qubit is not parity-protected---or ``topologically protected''---from decoherence~\cite{Footnote:decoherence}. Although this disqualifies Majorana modes of light for practical applications such as topological quantum memories, this crucially does \emph{not} hinder the observation of their exotic physics, since the key features of Majorana physics---such as the existence of localized Majorana modes and the possibility of simulating non-Abelian braiding operations in the associated zero-energy subspace---do not require perfect parity protection to be realized. In fact, as we demonstrate below, Majorana physics does remain accessible within time scales much shorter than the lifetime $\sim 1 / \Gamma$ of a photon in the system. However, the Majorana modes appearing in our optical framework only behave as genuine, \emph{local} Majorana modes in the limit $\Gamma \to 0$. Only then does the string-like operator $P$ that they carry---a remnant of the Jordan-Wigner mapping---reduce to a simple phase $P = \pm 1$. In this respect, we will refer to Majorana modes of light as ``Majorana-like'' modes (MLMs).

Our proposal for MLMs, embodied in Eqs.~\eqref{eqn:boseHubbard},~\eqref{eqn:driveHam}, and~\eqref{eqn:masterEquation}, could be realized in any cavity quantum electrodynamics (cavity QED) system lying deep in the strong-coupling regime. In the Supplemental Material~\cite{SupplementalMaterial}, we outline a potential implementation in circuit QED which we deem as closest to experimental realization, and give ballpark figures for the relevant parameters and energy scales.

\begin{figure}[t]
    \begin{center}
        \includegraphics[width=\columnwidth]{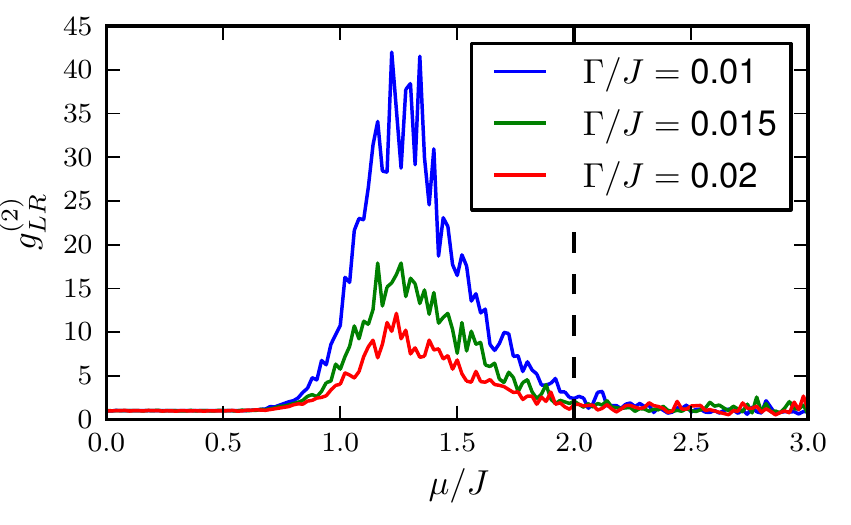}
        \caption{(Color online). Numerical results showing the second-order cross-correlation function $g^{(2)}_{LR}$ as a function of the detuning $\mu$ for $N = 12$ (including probe cavities), $\Delta / J = 1$, $J_{L,R} / J = 0.02$, and for different values of $\Gamma / J = \Gamma_{L,R} / J$. MLMs (with energy splitting $\delta_M$ determined by $\mu$) are expected in the region $0 \leq \mu / J < 2$ (delimited by a dashed vertical line) which corresponds to a topologically nontrivial phase (in the limit $N \to \infty$). As expected, a strong bunching signature is observed for large $\mu$ \emph{inside} the topologically nontrivial region, while $g^{(2)}_{LR} \approx 1$ beyond the critical point $\mu / J = 2$, clearly signaling the absence of MLMs. All results were obtained using a Monte Carlo wave-function approach (see, e.g.,~\cite{Yamamoto99}) with $400$ trajectories.}
        \label{fig:crossCorrelations}
    \end{center}
\end{figure}

\emph{Optical detection scheme.}---Multiple schemes have been proposed for detecting Majorana modes in solid-state systems (see, e.g.,~\cite{Alicea12} for a review). Although we believe that most of them can be transposed to our optical setting, we will focus on the detection of Majorana-mediated (photonic) Cooper pair splitting, following the proposals of Refs.~\cite{Nilsson08, Law09}. We start with a Kitaev chain in a topologically nontrivial regime defined by $J$, $\Delta > 0$, and $0 \leq \mu < 2J$, without loss of generality. In such a parameter range, exponentially localized MLMs are expected on both sides of the chain, with a length scale that increases with $\mu$ and diverges as $\mu \to 2J$~\cite{Kitaev01}. For finite $\mu < 2J$ and small enough system sizes, these modes weakly couple and the levels of the Majorana qubit that they form split in energy by an amount $\delta_M > 0$~\cite{Footnote:deltaM}, as captured by the Hamiltonian $\delta_M \sigma^{z}_M$ (the explicit form of $\sigma^{z}_M$ reduces to $\sigma^{y}_1 \sigma^{y}_N$ when restricted to the end cavities, and is given by Eq.~\eqref{eqn:sigmazM} for $\mu = 0$); as a key ingredient, we assume that $\delta_M \ll E_g$, where $E_g$ denotes the gap of the system~\cite{Footnote:gap}. Next we introduce two additional nonlinear cavities---one on each side of the Kitaev chain---which we refer to as the left (L) and right (R) probe cavities, respectively. We assume that the latter have resonance frequencies $\omega_{L,R} = \omega_p$, and that they couple to the end cavities of the Kitaev chain through weak tunneling only. In the rotating frame introduced above (in deriving Eq.~\eqref{eqn:kitaevHam}), the Hamiltonian describing the interaction with the probes then takes the form $H_{\text{probe}} = - J_L (\sigma^{x}_L \sigma^{x}_1 + \sigma^{y}_L \sigma^{y}_1) - J_R (\sigma^{x}_N \sigma^{x}_R + \sigma^{y}_N \sigma^{y}_R)$, where $0 < J_{L,R} \ll \delta_M$ denotes the weak amplitude for tunneling into the left and right probe cavities, respectively. Since all energy scales associated with $H_{\text{probe}}$ are, by assumption, much smaller than $E_g$, the probe cavities only probe the low-energy physics associated with the MLMs of the chain. Owing to this energy selectivity, the terms of the form $\sigma^{y}_i \sigma^{y}_j$ appearing in $H_{\text{probe}}$---which mediate coupling to higher excited states---can safely be neglected, while the operators $\sigma^{x}_1$ and $\sigma^{x}_N$---which anticommute with $\sigma^{z}_M$ and thus effectively describe a spin flip of the Majorana qubit---can be replaced by $\sigma^{x}_M$. This results in the following low-energy effective Hamiltonian for the full system:
\begin{align} \label{eqn:effProbeHam}
    H_{\text{eff}} & = \delta_M \sigma^{z}_M - J_L \sigma^{x}_L \sigma^{x}_M - J_R \sigma^{x}_M \sigma^{x}_R.
\end{align}
Physically, the above expression tells us that the nonlocal Majorana qubit formed by the ``localized'' (in the limit $\Gamma \to 0$) MLMs of the chain mediates a \emph{nonlocal} coherent exchange of photons between the probe cavities. Clearly, the bottleneck of such an exchange is given by the time scale $t_M \sim 1 / \delta_M$ over which the Majorana qubit evolves. We thus only expect to see nonlocal correlations between the probes if $t_M$ is the shortest time scale in Eq.~\eqref{eqn:effProbeHam}, i.e., if $\delta_M \gg J_{L,R}$, as has been assumed. To detect these correlations, one can take advantage of the intrinsic dissipative nature of the system. Assuming that the decay rate of the probe cavities satisfies $\Gamma_{L,R} \sim J_{L,R} \ll \delta_M$, so that spontaneous emission occurs on a time scale much longer than the time scale $t_M$ over which correlations are generated, we expect a direct signature of MLMs to appear in the second-order photon cross-correlations between the light emitted from the two probe cavities. In order to illustrate this, we consider the simple case $J_L = J_R \equiv \sqrt{2} \tilde{J}$, $\Gamma_L = \Gamma_R \equiv 8 \tilde{\Gamma}$, in the limit where the decay rate $\Gamma$ associated with the cavities of the Kitaev chain vanishes. Following the method of Ref.~\cite{Zunkovic10} (see Supplemental Material~\cite{SupplementalMaterial}), we then obtain steady-state photon cross-correlations between the probe cavities that read
\begin{align} \label{eqn:g2}
    g^{(2)}_{LR} & \equiv \frac{\avg{\tilde{b}^\dagger_L \tilde{b}^\dagger_R \tilde{b}_R \tilde{b}_L}}{\avg{\tilde{b}^\dagger_L \tilde{b}_L} \avg{\tilde{b}^\dagger_R \tilde{b}_R}} = 1 + \frac{\avg{\sigma^{z}_L \sigma^{z}_R} - \avg{\sigma^{z}_L} \avg{\sigma^{z}_R}}{(1 + \avg{\sigma^{z}_L}) (1 + \avg{\sigma^{z}_R})} \nonumber \\
    & = 1 + \frac{\tilde{\Gamma}^2 \delta^2_M}{(\tilde{J}^2 + \tilde{\Gamma}^2)^2}.
\end{align}
Remembering that $\tilde{\Gamma}^2 \sim \tilde{J}^2 \ll \delta^2_M$, we thus find $g^{(2)}_{LR} \gg 2$; in other words, the light emitted by the spatially separated probe cavities is strongly bunched~\cite{Footnote:g1}. To examine the effect of weak dissipation from the chain, we have carried out numerical simulations of the \emph{full} Kitaev chain coupled to the probe cavities. Our studies confirm that a striking nonlocal photon-bunching signature of MLMs remains visible for small decay rates $\Gamma \sim \Gamma_{L,R} \ll \delta_M$ of the chain cavities and small enough system sizes, i.e., as long as the effective width of the Majorana levels is much smaller than their energy splitting [Fig.~\ref{fig:crossCorrelations}].

We remark that the above results closely parallel those obtained in Refs.~\cite{Nilsson08, Law09} in the solid-state setting. Here, the probe cavities play a similar role as the metallic leads used in typical solid-state proposals: they provide a narrow band of states of effective width $\sim 1 / \Gamma_{L,R}$ close to the ``Fermi level'' $\omega_c$ into which fermionized photons can be emitted from the system. Due to the large splitting $\delta_M$ between MLMs, nonlocal---or ``crossed''---Andreev reflection is favored over its local analog, and Cooper pairs of photons are split into separate leads (or probe cavities). We remark that one could in principle remove the probe cavities and observe directly the light emitted by the end cavities of the Kitaev chain. In that case, however, spectral filtering would be required in order to isolate the bunching signature associated with the low-energy MLMs from contributions of the bulk.


\emph{Versatility of the optical proposal beyond detection.}---In contrast to previous proposals (such as Refs.~\cite{Oreg10, Lutchyn10, Sau12}), our optical system provides a very versatile platform for simulating Majorana physics. In addition to being a conceptually simple realization of Kitaev's original lattice toy model---fermionized photons are intrinsically spinless, and pairing occurs between nearest-neighboring cavities only---it allows for single-site addressability and for local control of all parameters entering Kitaev's model: the ``chemical potential'' $\mu$ can be tuned easily and locally by modifying the frequency of the individual pumps (or, alternatively, the resonance frequency of the individual cavities), and the amplitude and phase of the ``superconducting order parameter'' $\Delta$ can similarly be adjusted by controlling the amplitude and phase of the parametric amplifiers. The tunneling amplitude $J$, on the other hand, can be tuned by introducing intermediate control devices between cavities (see, e.g., Ref.~\cite{Houck12} and refs. therein). Such level of control is key to overcoming the crucial challenges currently facing most solid-state proposals, such as the tuning in and out of the topological phase and the suppression of disorder effects~\cite{Sau12}.

Despite its conceptual simplicity, versatility, and physical realizability, our optical proposal departs from being ideal in two respects: (i) it lacks topological protection (or parity conservation), and (ii) it cannot be scaled up to networks of 1D wires~\cite{Alicea11, Halperin12}. The first imperfection arises as a direct consequence of photon losses---unavoidable in photonic systems---and puts stringent constraints on the time scale over which Majorana physics can be observed; namely, MLMs must be manipulated and detected on a time scale much shorter than the lifetime of a photon in the system. We argue in the Supplemental Material~\cite{SupplementalMaterial}, however, that state-of-the-art technologies in circuit QED could allow for the experimental realization of our proposal with a sufficient control over dissipation to meet these requirements. The second imperfection stems from the intrinsic nonlocal nature of the Jordan-Wigner mapping invoked in deriving Eq.~\eqref{eqn:kitaevHam}, ruling out the possibility to observe non-Abelian exchange statistics in connected wire geometries. In 1D, the Jordan-Wigner string carried by the end MLMs of a chain essentially corresponds to the parity operator of the latter (see, e.g., Eq.~\eqref{eqn:sigmazM}), such that end MLMs do behave as genuine, local Majorana modes \emph{on time scales over which parity is effectively constant}. In higher dimensions, however, the situation changes drastically: when multiple 1D chains are contacted (not through their ends, so that the systems effectively is higher-dimensional), the parity of the individual chains becomes a dynamical quantity and the nonlocal nature of the MLMs comes into play---with dramatic consequences such as the absence of non-Abelian exchange statistics. To avoid such complications, we simply strictly restrict ourselves to 1D systems. In that case, the exchange of Majorana modes is impossible in real space, but can nevertheless be \emph{simulated} using so-called ``tunnel-braid'' operations~\cite{Flensberg11}. As shown in the Supplemental Material~\cite{SupplementalMaterial}, these operations only preserve the degeneracy of the MLMs---as real-space braiding---provided that the relative phase between the latter is properly tuned, and therefore are not strictly speaking topologically protected. This, however, does not constitute an additional problem in our optical setting where parity is anyway not conserved. In the framework of our proposal, tunnel-braid operations crucially allow us to obviate the need for real-space braiding, hence providing us with a full-fledged 1D optical platform for Majorana physics.


\emph{Conclusions.}---We anticipate that our proposal for realizing and detecting photonic $p$-wave pairing will allow for an exciting alternative avenue for the investigation of Majorana physics. An interesting possibility would be the investigation of Majorana modes in a continuum 1D model of strongly interacting optical photons~\cite{Chang08}.


The authors would like to thank Andreas Wallraff and Christopher Eichler for insightful discussions regarding the possible implementations in circuit QED, and Jeroen Elzerman, Ajit Srivastava and Bj\"orn Sbierski for critical comments. This work was supported by an ERC Advanced Investigator Grant, and by NCCR Quantum Science and Technology (NCCR QSIT), research instrument of the Swiss National Science Foundation (SNSF).



\begin{thebibliography}{99}


\bibitem{Brune96}
M. Brune, F. Schmidt-Kaler, A. Maali, J. Dreyer, E. Hagley, J. M. Raimond, and S. Haroche, Phys. Rev. Lett. {\bf 76}, 1800 (1996).

\bibitem{Fink08}
J. M. Fink, M. G\"oppl, M. Baur, R. Bianchetti, P. J. Leek, A. Blais, and A. Wallraff, Nature {\bf 454}, 315 (2008).

\bibitem{Imamoglu97}
A. \.Imamo\u glu, H. Schmidt, G. Woods, and M. Deutsch, Phys. Rev. Lett. {\bf 79}, 1467 (1997).

\bibitem{Birnbaum05}
K. M. Birnbaum, A. Boca, R. Miller, A. D. Boozer, T. E. Northup, and H. J. Kimble, Nature Phys. {\bf 436}, 87 (2005).

\bibitem{Faraon08}
A. Faraon, I. Fushman, D. Englund, N. Stoltz, P. Petroff, and J. Vuckovic, Nature Phys. {\bf 4}, 859 (2008).

\bibitem{Bozyigit11}
D. Bozyigit, C. Lang, L. Steffen, J. M. Fink, C. Eichler, M. Baur, R. Bianchetti, P. J. Leek, S. Filipp, M. P. da Silva, A. Blais, and A. Wallraff, Nature Phys. {\bf 7}, 154 (2011).

\bibitem{Hoffman11}
A. J. Hoffman, S. J. Srinivasan, S. Schmidt, L. Spietz, J. Aumentado, H. E. T\"ureci, and A. A. Houck, Phys. Rev. Lett. {\bf 107}, 053602 (2011).

\bibitem{Reinhard12}
A. Reinhard, T. Volz, M. Winger, A. Badolato, K. J. Hennessy, E. L. Hu, and A. \.Imamo\u glu, Nature Photon. {\bf 6}, 93 (2012).

\bibitem{Reich12}
E. S. Reich, Nature News {\bf 483}, 132 (2012), doi:10.1038/ 483132a.

\bibitem{Alicea12}
J. Alicea, arXiv:1202.1293 (2012).

\bibitem{Read00}
N. Read and D. Green, Phys. Rev. B {\bf 61}, 10267 (2000).

\bibitem{Chang08}
D. E. Chang, V. Gritsev, G. Morigi, V. Vuletic, M. D. Lukin, E. A. Demler, Nature Phys. {\bf 4}, 884 (2008).

\bibitem{Carusotto09}
I. Carusotto, D. Gerace, H. E. T\"ureci, S. De Liberato, C. Ciuti, and A. \.Imamo\u glu, Phys. Rev. Lett. {\bf 103}, 033601 (2009).

\bibitem{Kitaev01}
A. Yu. Kitaev, Phys.-Usp. {\bf 44}, 131 (2001).

\bibitem{Trauzettel11}
J. C. Budich, S. Walter, and B. Trauzettel, arXiv: 1111.1734 (2011).

\bibitem{Bardyn12}
C.-E. Bardyn, M. A. Baranov, E. Rico, A. \.Imamo\u glu, P. Zoller, and S. Diehl, arXiv:1201.2112 (2012).

\bibitem{Kitaev03}
A. Yu. Kitaev, Ann. Phys. {\bf 303}, 2 (2003).

\bibitem{Flensberg11}
K. Flensberg, Phys. Rev. Lett. {\bf 106}, 090503 (2011).

\bibitem{Hartmann06}
M. J. Hartmann, F. G. S. L. Brandao, and M. B. Plenio, Nature Phys. {\bf{2}}, 849 (2006).

\bibitem{Footnote:rotatingFrame}
$\omega_c > U$ is not relevant in the rotating frame used later in the text.

\bibitem{SupplementalMaterial}
See online Supplemental Material.

\bibitem{Footnote:pWavePairing}
The $p$-wave nature of the pairing will be made clear in the fermionic picture used later in the text.

\bibitem{Cazalilla11}
M. A. Cazalilla, R. Citro, T. Giamarchi, E. Orignac, and M. Rigol, Rev. Mod. Phys. {\bf 83}, 1405 (2011).

\bibitem{Footnote:decoherence}
Decoherence increases with the length of the string corresponding to $P$, i.e. with the system size.

\bibitem{Nilsson08}
J. Nilsson, A. R. Akhmerov, and C. W. J. Beenakker, Phys. Rev. Lett. {\bf 101}, 120403 (2008).

\bibitem{Law09}
K. T. Law, P. A. Lee, and T. K. Ng, Phys. Rev. Lett. {\bf 103}, 237001 (2009).

\bibitem{Footnote:deltaM}
$\delta_M \sim e^{-N / \xi}$ with $\xi^{-1} = \min{ \{ \abs{\ln{\abs{x_{+}}}}, \abs{\ln{\abs{x_{-}}}} \} }$ and $x_{\pm} = \tfrac{-\mu \pm \sqrt{\mu^2 - 4J^2 + 4\abs{\Delta}^2}}{2(J + \abs{\Delta})}$~\cite{Kitaev01}.

\bibitem{Footnote:gap}
In the parameter regime of interest, $E_g = 2J - \mu$ if $\Delta \geq J$ or if $\Delta < J$ and $2J - \mu < 2 \Delta^2 / J$, and $E_g = \Delta (4 - \mu^2 / (J^2 - \Delta^2))^{1/2}$ otherwise~\cite{Kitaev01}.

\bibitem{Zunkovic10}
B. \v{Z}unkovi\v{c} and T. Prosen, J. Stat. Mech. ({\bf 2010}) P08016.

\bibitem{Footnote:g1}
The steady-state population of the probe cavities scales as $1 / g^{(2)}_{LR}$ (see Supplemental Material~\cite{SupplementalMaterial}). In practise, we thus want to work in a regime where nonlocal bunching is not too strong, so that the occupation of the probe cavities allows for a good detection efficiency.

\bibitem{Yamamoto99}
Y. Yamamoto and A. \.Imamo\u glu, \emph{Mesoscopic Quantum Optics} (Wiley, New York, 1999).

\bibitem{Oreg10}
Y. Oreg, G. Refael, and F. von Oppen, Phys. Rev. Lett. {\bf 105}, 177002 (2010).

\bibitem{Lutchyn10}
R. M. Lutchyn, J. D. Sau, and S. Das Sarma, Phys. Rev. Lett. {\bf 105}, 077001 (2010).

\bibitem{Sau12}
J. D. Sau and S. Das Sarma, Nature Commun. {\bf 3}, 964 (2012).

\bibitem{Houck12}
A. A. Houck, H. E. T\"ureci, and J. Koch, Nature Phys. {\bf 8}, 292 (2012).

\bibitem{Alicea11}
J. Alicea, Y. Oreg, G. Refael, F. von Oppen, and M. P. A. Fisher, Nature Phys. {\bf 7}, 412 (2011).

\bibitem{Halperin12}
B. I. Halperin, Y. Oreg, A. Stern, G. Refael, J. Alicea, and F. von Oppen, Phys. Rev. B {\bf 85}, 144501 (2012).


\end{thebibliography}
\end{document}



\title{Supplemental Material for ``Majorana-Like Modes of Light in a One-Dimensional Array of Nonlinear Cavities''}

\author{C.-E. Bardyn}
\affiliation{Institute for Quantum Electronics, ETH Zurich, 8093 Zurich, Switzerland}

\author{A. \.Imamo\u glu}
\affiliation{Institute for Quantum Electronics, ETH Zurich, 8093 Zurich, Switzerland}

\maketitle


\section{$P$-wave pairing of strongly correlated photons}

In this section, we demonstrate explicitly that $p$-wave pairing generally emerges as a result of the interplay between two-photon parametric pumping and strong on-site photon-photon repulsion, and derive the effective drive Hamiltonian of Eq.~(2) in the main text. Although multiple schemes can be envisioned to realize the key nonlinear pumping process that takes photons from a pump field and injects coherent photon pairs into the system, we will illustrate the general mechanism that leads to $p$-wave pairing in a model based on a theoretical description of coupled cavity arrays that was developed in Ref.~\cite{Hartmann06}.

We start by assuming that the parametric pumps act locally in between the nearest-neighboring cavities corresponding to the sites $i$ and $i+1$ according to the standard Hamiltonian~\cite{Hong85}
\begin{align} \label{eqn:driveHam}
    H_{\text{drive}, i} & = \int \mathrm{d}^3 r \, \chi^{(2)} (\br) E^{(+)}_{p, i} (\br, t) E^{(-)}_{s, i} (\br) E^{(-)}_{s, i} (\br) + h.c.,
\end{align}
where $E^{(+)}_{p, i} (\br, t)$ is the positive-frequency part of the pump (index ``p'') optical field, $E^{(-)}_{s, i} (\br)$ the negative-frequency part of the generated signal (index ``s'') optical fields, and $\chi^{(2)} (\br)$ the effective second-order optical nonlinearity of the system (taking into account the polarization vectors of the three interacting fields). We assume that the pump field can be treated as a classical monochromatic field of frequency $2 \omega_{p, i}$ and complex amplitude $E^0_{p, i}$, and that there is no significant depletion of the latter such that $E^0_{p, i}$ can be considered as constant. Denoting the spatial mode function of the pump as $\varphi_{p, i} (\br)$, $E^{(+)}_{p, i} (\br, t)$ then takes the form
\begin{align} \label{eqn:pumpField}
    E^{(+)}_{p, i} (\br, t) & = E^0_{p, i} \varphi_{p, i} (\br) e^{-2 \mathrm{i} \omega_p t}.
\end{align}
The generated signal optical fields, on the other hand, can generally be expanded in terms of the Wannier modes $\phi_i (\br)$ of the coupled cavity array. Assuming that these modes decay sufficiently fast so that only nearest-neighboring modes have a non-vanishing overlap, we can write
\begin{align} \label{eqn:signalField}
    E^{(-)}_{s, i} (\br) & = -\mathrm{i} \, \sqrt{\omega_c} \left( \phi^*_i (\br) b^\dagger_i + \phi^*_{i+1} (\br) b^\dagger_{i+1} \right),
\end{align}
where $\omega_c$ is the resonance frequency of the cavity (assumed to be the same for all cavities). Plugging Eqs.~\eqref{eqn:pumpField} and~\eqref{eqn:signalField} into Eq.~\eqref{eqn:driveHam}, we thus obtain
\begin{align}
    H_{\text{drive}, i} & = -E^0_{p, i} \, \omega_c \int \mathrm{d}^3 r \, \chi^{(2)} (\br) \varphi_{p, i} (\br) \left( \phi^{* 2}_i (\br) b^{\dagger 2}_i + \phi^*_i (\br) \phi^*_{i+1} (\br) b^\dagger_i b^\dagger_{i+1} + \phi^{* 2}_{i+1} (\br) b^{\dagger 2}_{i+1} \right) e^{-2 \mathrm{i} \omega_p t} + h.c. \nonumber \\
    & \approx -E^0_{p, i} \, \omega_c \left( \int \mathrm{d}^3 r \, \chi^{(2)} (\br) \varphi_{p, i} (\br) \phi^*_i (\br) \phi^*_{i+1} (\br) \right) b^\dagger_i b^\dagger_{i+1} e^{-2 \mathrm{i} \omega_p t} + h.c.,
\end{align}
where we have invoked the presumably strong on-site repulsion between photons to neglect all terms of the form $b^{\dagger 2}_i$. The strength of the degenerate parametric pumping therefore clearly appears to depend on the overlap of the interacting modes, as expected. Assuming, for simplicity, that the pump field and the effective second-order optical nonlinearity are essentially constant over the volume $V_i$ between the cavities, and that the pump field vanishes outside this volume, we finally obtain
\begin{align}
    H_{\text{drive}, i} & \approx \kappa_i \left( \int_{V_i} \mathrm{d}^3 r \, \phi^*_i (\br) \phi^*_{i+1} (\br) \right) b^\dagger_i b^\dagger_{i+1} e^{-2 \mathrm{i} \omega_p t} + h.c. \nonumber \\
    & \equiv - \Delta^*_i \, b^\dagger_i b^\dagger_{i+1} e^{-2 \mathrm{i} \omega_p t} + h.c.,
\end{align}
where the complex coupling constant $\kappa_i$ essentially depends (linearly) on $\chi^{(2)}$ and on the amplitude of the pump field $E^0_{p, i}$. This last result shows that the strength of the two-photon parametric drive (i.e., the $p$-wave pairing strength) directly depends on the overlap of the Wannier modes corresponding to nearest-neighboring cavities, as argued in the main text.


\section{Steady-state second-order photon cross-correlations}

In what follows, we give a brief derivation of Eq.~(11) from the main text. In accordance with the discussion presented there, we start with the Lindblad master equation
\begin{align} \label{eqn:masterEquation}
    \partial_t \rho & = - \mathrm{i} \left[ H_{\text{eff}}, \rho \right] + \sum_{i=L,R} \left( L_i \rho \, L^\dagger_i - \tfrac{1}{2} \{ L^\dagger_i L_i, \rho \} \right),
\end{align}
where the Hamiltonian and the Lindblad operators are respectively given by
\begin{align}
    H_{\text{eff}} & = \delta_M \sigma^{z}_M - \sqrt{2} \tilde{J} (\sigma^{x}_L \sigma^{x}_M + \sigma^{x}_M \sigma^{x}_R), \\
    L_i & = \sqrt{\tfrac{\tilde{\Gamma}}{2}} \sigma^{-}_i.
\end{align}
Regarding the system as a chain of $3$ interacting spins and relabeling the sites as $L \rightarrow 1$, $M \rightarrow 2$, $R \rightarrow 3$, one can perform a Jordan-Wigner transformation and describe the problem in a basis of Majorana operators defined as
\begin{align}
    \setlength{\extrarowheight}{3pt}
    \begin{array}{ll}
        c_{2i-1} & = (\prod_{j=1}^{i-1} \sigma^{z}_j) \sigma^{x}_i, \\
        c_{2i} & = (\prod_{j=1}^{i-1} \sigma^{z}_j) \sigma^{y}_i.
    \end{array}
\end{align}
In this fermionic description, the Hamiltonian and the Lindblad operators take the form
\begin{align}
    H_{\text{eff}} & = - \mathrm{i} \delta_M c_3 c_4 - \mathrm{i} \sqrt{2} \tilde{J} (c_2 c_3 + c_4 c_5), \\
    L_1 & = \sqrt{\tfrac{\tilde{\Gamma}}{2}} (c_1 - \mathrm{i} c_2), \\
    L_3 & = \sqrt{\tfrac{\tilde{\Gamma}}{2}} (c_5 - \mathrm{i} c_6),
\end{align}
and can conveniently be expressed as $H_{\text{eff}} = \tfrac{\mathrm{i}}{2} \vect{c}^T H \vect{c}$ and $L_i = \vect{l}_{i}^T \vect{c}$, where $\vect{c}^T = (c_1, c_2, \ldots, c_{6})$. Note that we have dropped the string-like operator of the form $\prod_{j=1}^{3} \sigma^{z}_j$ that would normally appear in $L_3$, since the latter does not affect Eq.~\eqref{eqn:masterEquation} (see, e.g., Ref.~\cite{Zunkovic10}). The fact that Eq.~\eqref{eqn:masterEquation} is quadratic allows for a great simplification of the problem. Assuming, without loss of generality, that the initial state of the system has a Gaussian form, one can focus exclusively on the time evolution of the correlation matrix $C_{i, j} = (\mathrm{i} / 2) \operatorname{tr} ([ c_i, c_j ] \rho)$ which contains all information about the system. As shown in Ref.~\cite{Eisert10}, the master equation of Eq.~\eqref{eqn:masterEquation} then reduces to a matrix equation of the form
\begin{align}
    \partial_t C = X^T C + C X - Y,
\end{align}
where $X = - 2 \mathrm{i} H + 2 \operatorname{Re} M$ and $Y = 4 \operatorname{Im} M$, with $M = \sum_i \vect{l}_{i} \otimes \vect{l}^\dagger_i$. Here we find
\begin{align}
    X & =
    \left(
    \begin{array}{cccccc}
        \tilde{\Gamma}  & 0 & 0 & 0 & 0 & 0 \\
        0 & \tilde{\Gamma}  & 2 \sqrt{2} \tilde{J} & 0 & 0 & 0 \\
        0 & -2 \sqrt{2} \tilde{J} & 0 & -\delta_M & 0 & 0 \\
        0 & 0 & \delta_M & 0 & 2 \sqrt{2} \tilde{J} & 0 \\
        0 & 0 & 0 & -2 \sqrt{2} \tilde{J} & \tilde{\Gamma}  & 0 \\
        0 & 0 & 0 & 0 & 0 & \tilde{\Gamma}  \\
    \end{array}
    \right), \quad
    Y & =
    \left(
    \begin{array}{cccccc}
        0 & 2 \tilde{\Gamma}  & 0 & 0 & 0 & 0 \\
        -2 \tilde{\Gamma}  & 0 & 0 & 0 & 0 & 0 \\
        0 & 0 & 0 & 0 & 0 & 0 \\
        0 & 0 & 0 & 0 & 0 & 0 \\
        0 & 0 & 0 & 0 & 0 & 2 \tilde{\Gamma}  \\
        0 & 0 & 0 & 0 & -2 \tilde{\Gamma}  & 0 \\
    \end{array}
    \right),
\end{align}
and the steady-state correlation matrix satisfies
\begin{align} \label{eqn:steadyState}
    X^T C + C X = Y.
\end{align}
In order to solve the above matrix equation, we follow the method of Ref.~\cite{Zunkovic10} and make the ansatz
\begin{align}
    X = \sum_{j = 1}^{6} \alpha_{j, k} (\vect{w}^T_j \otimes \vect{w}_k),
\end{align}
where $\vect{v}_j$ and $\vect{w}_j$ are left and right eigenvectors of $X$, respectively. Introducing this ansatz into Eq.~\eqref{eqn:steadyState} and assuming that $\vect{v}_j$, $\vect{w}_j$ are normalized according to $\vect{v}^\dagger_j \vect{v}_k = \delta_{j, k}$, we find
\begin{align}
    \alpha_{j, k} = \frac{1}{\lambda_j + \lambda_k} (\vect{v}^\dagger_j Y \vect{v}^*_k),
\end{align}
where $\lambda_j$ are the eigenvalues corresponding to the right (or left) eigenvectors of $X$. Therefore, solving for the steady-state correlation matrix basically amounts to solving two eigenvalue problems: $X \vect{w}_j = \lambda_j \vect{w}_j$ and $\vect{v}_j X = \lambda_j \vect{v}_j$. Using the above explicit form of $X$ and $Y$, straightforward algebra leads to
\begin{align}
    C & =
    \left(
    \begin{array}{cccccc}
        0 & \frac{\gamma ^2 \left(\gamma ^2+t^2+1\right)}{\gamma ^2+\left(\gamma ^2+t^2\right)^2} &
        \frac{\sqrt{2} \gamma  t \left(\gamma ^2+t^2\right)}{\gamma ^2+\left(\gamma ^2+t^2\right)^2} &
        -\frac{\sqrt{2} \gamma ^2 t}{\gamma ^2+\left(\gamma ^2+t^2\right)^2} & -\frac{\gamma  t^2}{\gamma
        ^2+\left(\gamma ^2+t^2\right)^2} & 0 \\
        -\frac{\gamma ^2 \left(\gamma ^2+t^2+1\right)}{\gamma ^2+\left(\gamma ^2+t^2\right)^2} & 0 & 0 & 0 & 0
        & \frac{\gamma  t^2}{\gamma ^2+\left(\gamma ^2+t^2\right)^2} \\
        -\frac{\sqrt{2} \gamma  t \left(\gamma ^2+t^2\right)}{\gamma ^2+\left(\gamma ^2+t^2\right)^2} & 0 & 0 &
        0 & 0 & -\frac{\sqrt{2} \gamma ^2 t}{\gamma ^2+\left(\gamma ^2+t^2\right)^2} \\
        \frac{\sqrt{2} \gamma ^2 t}{\gamma ^2+\left(\gamma ^2+t^2\right)^2} & 0 & 0 & 0 & 0 & -\frac{\sqrt{2}
        \gamma  t \left(\gamma ^2+t^2\right)}{\gamma ^2+\left(\gamma ^2+t^2\right)^2} \\
        \frac{\gamma  t^2}{\gamma ^2+\left(\gamma ^2+t^2\right)^2} & 0 & 0 & 0 & 0 & \frac{\gamma ^2
        \left(\gamma ^2+t^2+1\right)}{\gamma ^2+\left(\gamma ^2+t^2\right)^2} \\
        0 & -\frac{\gamma  t^2}{\gamma ^2+\left(\gamma ^2+t^2\right)^2} & \frac{\sqrt{2} \gamma ^2 t}{\gamma
        ^2+\left(\gamma ^2+t^2\right)^2} & \frac{\sqrt{2} \gamma  t \left(\gamma ^2+t^2\right)}{\gamma
        ^2+\left(\gamma ^2+t^2\right)^2} & -\frac{\gamma ^2 \left(\gamma ^2+t^2+1\right)}{\gamma
        ^2+\left(\gamma ^2+t^2\right)^2} & 0 \\
    \end{array}
    \right),
\end{align}
where we have defined the dimensionless quantities $t = \tilde{J} / \delta_M$, $\gamma = \tilde{\Gamma} / \delta_M$. The steady-state second-order photon cross-correlations between the left and right probes then read
\begin{align}
    g^{(2)}_{LR} & = 1 + \frac{\avg{\sigma^{z}_L \sigma^{z}_R} - \avg{\sigma^{z}_L} \avg{\sigma^{z}_R}}{(1 + \avg{\sigma^{z}_L}) (1 + \avg{\sigma^{z}_R})} \nonumber \\
    & = 1 + \frac{\avg{c_1 c_5} \avg{c_2 c_6} - \avg{c_1 c_6} \avg{c_2 c_5}}{(1 - \mathrm{i} \avg{c_1 c_2}) (1 -\mathrm{i} \avg{c_5 c_6})} \nonumber \\
    & = 1 - \frac{C_{1, 5} C_{2, 6} - C_{1, 6} C_{2, 5}}{(1 - C_{1, 2}) (1 - C_{5, 6})} \nonumber \\
    & = 1 + \frac{\gamma^2}{(t^2 + \gamma^2)^2},
\end{align}
where we have used Wick's theorem in the second equality. Similarly, the steady-state occupation of the probe cavities is given by
\begin{align}
    1 + \avg{\sigma^{z}_{L,R}} & = \frac{t^2 (t^2 + \gamma^2)}{\gamma^2 + (t^2 + \gamma^2)^2} \nonumber \\
    & = \frac{t^2}{t^2 + \gamma^2} \frac{1}{g^{(2)}_{LR}},
\end{align}
and scales as $1/g^{(2)}_{LR}$ when $\tilde{J} \sim \tilde{\Gamma}$.

\section{Implementation in circuit QED}

Our proposal for accessing Majorana physics in the optical context requires two crucial ingredients: strong coupling and (comparatively) weak photon losses, both of which are currently available in superconducting-circuit-based cavity quantum electrodynamics (circuit QED) (see Ref.~\cite{Houck12}, e.g., for a recent review). Superconducting qubits thus stand apart as an ideal platform for realizing our proposal---even more so because of the great flexibility in fabrication and control provided by state-of-the-art technologies. We outline below a potential implementation of our proposal which we deem as closest to experimental realization, and give ballpark figures for the relevant parameters and energy scales.

The system that we typically envision consists of a chain of capacitively-coupled identical microwave resonators (playing the role of cavities), each incorporating two superconducting transmon qubits (playing the role of artificial atoms). The first of these qubits is positioned at an antinode of the intracavity field, with a cavity-qubit coupling strength $g \simeq \omega_b \sqrt{\alpha}$, where $\omega_b$ is the bare-cavity resonance frequency and $\alpha$ the fine-structure constant; the strength of the effective on-site interaction---or, equivalently, of the anharmonicity---associated with the lower polariton mode (with frequency $\omega_c = \omega_b - g$) is then given by $U = (2 - \sqrt{2}) g$. The second qubit is located in between pairs of neighboring cavities, and is driven parametrically using strong microwave fields at frequency $2 \omega_p$ so as to ensure the generation or annihilation of photon pairs at frequency $\omega_b - g$. In the limit where the amplitude of the parametric drive (or ``pump'') is weak as compared to $g$, the only energetically allowed process is the generation (annihilation) of photon pairs in neighboring cavities, as discussed in the main text and in the first section above. This results in an effective $p$-wave pairing interaction with a phase determined by that of the microwave drive fields.

We emphasize that $g \sim 0.1 \omega_b \sim 10^4 \Gamma$, where $\Gamma$ is the cavity decay rate (see main text), is routinely obtained in state-of-the-art circuit-QED devices. This implies that the tunneling amplitude $J$ between neighboring cavities and the amplitude $\abs{\Delta}$ of the parametric pumps can in principle be tailored to satisfy the relation $U \sim g \gg J, \abs{\Delta} \gg \Gamma$ which lies at the core of our proposal. The first inequality ($U \sim g \gg J, \abs{\Delta}$) ensures that the system lies deep in the strong-coupling regime where photons effectively behave as fermions (or ``fermionized'' photons), and constitutes a sufficient condition for our model to map to that of Kitaev (see main text). If one additionally adjust the detuning $\mu = \omega_p - \omega_c$ between the cavities and the pumps so that $\abs{\mu} < 2J$, the system is found in a topologically nontrivial regime and exhibits Majorana-like modes with an energy splitting $\delta_M$ that depends on $J$, $\abs{\Delta}$, and $\abs{\mu}$, and vanishes exponentially with the system size $N$ (see the explicit form below). The second inequality ($J, \abs{\Delta} \gg \Gamma$), in contrast, determines the \emph{broadening} of the Majorana levels, i.e., how well they can be resolved for a fixed energy splitting $\delta_M$. For Majorana-like modes to be detectable using the scheme presented in the main text, this broadening must crucially be (i) much smaller than the energy splitting $\delta_M$ between the Majorana levels, and (ii) much smaller than the gap $E_g$ of the system, so that Majorana levels do not overlap in energy with levels of the bulk. In general, these two conditions depend on the system parameters in a nontrivial way given by the explicit form of $\delta_M$ and $E_g$~\cite{Kitaev01}, namely,
\begin{itemize}
    \item $\delta_M \sim e^{-N / \xi}$, where $\xi$ is the localisation length of the Majorana-like modes given by $\xi^{-1} = \min{ \{ \abs{\ln{\abs{x_{+}}}}, \abs{\ln{\abs{x_{-}}}} \} }$ with $x_{\pm} = \tfrac{-\mu \pm \sqrt{\mu^2 - 4J^2 + 4\abs{\Delta}^2}}{2(J + \abs{\Delta})}$,
    \item $E_g = 2J - \abs{\mu}$ if $\abs{\Delta} \geq J$ or if $\abs{\Delta} < J$ and $2J - \abs{\mu} < 2 \abs{\Delta}^2 / J$, and $E_g = \abs{\Delta} (4 - \abs{\mu}^2 / (J^2 - \abs{\Delta}^2))^{1/2}$ otherwise (in the topologically nontrivial regime with $2J - \abs{\mu} > 0$).
\end{itemize}
Away from the limit $2J - \abs{\mu} \to 0$ where the gap closes (and where the topological phase transition occurs, in the limit $N \to \infty$), $\delta_M \gg \Gamma$ is generally more restrictive than $E_g \gg \Gamma$. As long as the above inequality $J, \abs{\Delta} \gg \Gamma$ is ensured, both of these conditions can always be satisfied by tuning $\mu$ well inside the topologically nontrivial region, as shown in the numerical results presented in the main text. $J, \abs{\Delta} \gg \Gamma$ is therefore a \emph{sufficient} condition for our detection scheme to work, \emph{up to the tuning of $\mu$}. This condition is also necessary, since $J, \abs{\Delta} \sim \Gamma$ would make the broadening comparable to the gap.

Although this is not a requirement for our proposal, we finally remark that one typically expects to be able to work in a regime where $J \sim \abs{\Delta}$, since both quantities are essentially determined by the overlap of the Wannier modes of neighboring cavities. We also emphasize that the fabrication of a linear chain of $\sim 10$ almost identical cavities is technologically feasible, as suggested by the recent realization of a Kagome lattice of more than $200$ nearly identical cavities~\cite{Underwood12, Houck12}. Photon correlation measurements on microwave cavities have also been recently demonstrated~\cite{Bozyigit11}.

\section{Simulating real-space braiding in 1D using tunnel-braid operations}

The ability to exchange---or ``braid''---Majorana modes in real space plays an important role in demonstrating the non-Abelian nature of the latter. In the framework of the optical proposal presented in the main text, however, one is fundamentally restricted to one-dimensional systems where exchanging Majorana modes is physically impossible (Majorana modes would overlap and split in energy during the exchange, thereby hybridizing into complex fermionic modes and losing their exotic nature). It thus appears crucial to find another way to perform non-Abelian operations in the degenerate ground-state subspace associated with Majorana modes \emph{while preserving their degeneracy}. Such a possibility was originally introduced in Ref.~\cite{Flensberg11}, where it was shown that so-called ``tunnel-braid'' operations can \emph{simulate} the braiding of Majorana modes in real space. In what follows, we translate these ideas into the framework of our optical proposal, and give an explicit procedure to simulate the exchange of two Majorana-like modes of light tunnel-coupled to the same ancillary cavity.

\begin{figure}[h]
    \begin{center}
        \includegraphics{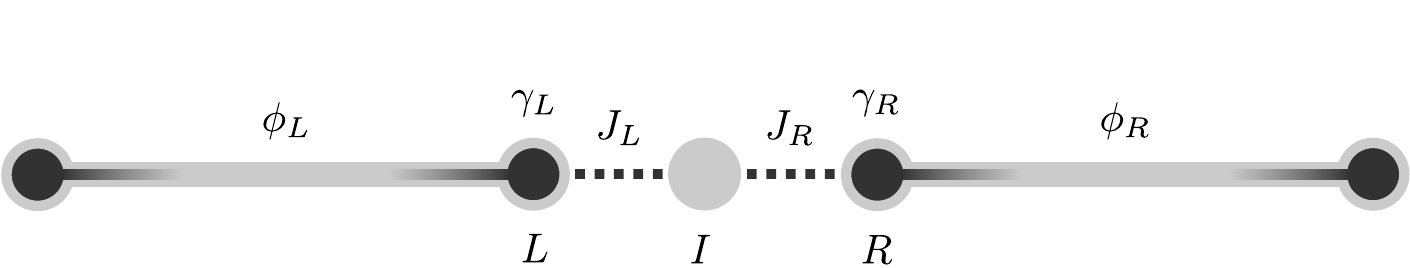}
        \caption{Basic one-dimensional setup required to simulate non-Abelian operations such as braiding. Two Kitaev chains---left (L) and right (R)---with pump phases $\phi_L$ and $\phi_R$, respectively, exhibit Majorana-like modes $\gamma_L$ and $\gamma_R$ at their ends, as depicted. Each chain tunnel-couples with amplitude $J_L$ ($J_R$) to an intermediate nonlinear cavity (I) via its end cavity where $\gamma_L$ ($\gamma_R$) is mostly localized.}
        \label{fig:tunnelBraidSetup}
    \end{center}
\end{figure}

To illustrate the working principles of tunnel-braid operations, we consider a simple scenario where two Kitaev chains---left (L) and right (R)---are end-to-end tunnel-coupled through a single intermediate cavity (I) [Fig.~\ref{fig:tunnelBraidSetup}]. We assume that the parametric pumps driving both chains have phases $\phi_L$ and $\phi_R$, respectively, and the same frequency $\omega_p$. The intermediate cavity is nonlinear---though not driven---and resonant with $\omega_p$, and both chains are in a topologically non-trivial phase with Majorana-like zero modes $\gamma_L$ and $\gamma_R$ located on the left and on the right of the intermediate cavity, respectively. Photon tunneling occurs between the intermediate cavity and the end cavities of the Kitaev chains, as captured by the Hamiltonian
\begin{align}
    H_{\text{t}} & = - 2 J_L (b^\dagger_L b_I + h.c.) - 2 J_R (b^\dagger_I b_R + h.c.),
\end{align}
where $b_L$, $b_I$ and $b_R$ are annihilation operators associated with photons in the respective cavities (with obvious notations), while $J_L$ and $J_R$ are positive amplitudes for tunneling into the left and right chains, respectively (factors of $2$ are introduced for later convenience). Transforming to the fermionic picture defined in the main text, we obtain
\begin{align} \label{eqn:tunnelHam}
    H_{\text{t}} & = - 2 J_L (a^\dagger_L a_I + h.c.) - 2 J_R (a^\dagger_I a_R + h.c.),
\end{align}
with fermionic operators $a_L$ and $a_R$ that can be decomposed as
\begin{align}
    \begin{split}
        a_L & = \tfrac{1}{2} e^{- \mathrm{i} \phi_L / 2} (\gamma'_L + \mathrm{i} \gamma_L), \\
        a_R & = \tfrac{1}{2} e^{- \mathrm{i} \phi_R / 2} (\gamma_R + \mathrm{i} \gamma'_R),
    \end{split}
\end{align}
where $\gamma'_L$ and $\gamma'_R$ are Majorana operators whose explicit form is not relevant here. Since we are only interested in the low-energy physics associated with the Majorana-like zero modes $\gamma_L$ and $\gamma_R$, we can make in Eq.~\eqref{eqn:tunnelHam} the following replacement:
\begin{align}
    \begin{split}
        a_L & \rightarrow \tfrac{\mathrm{i}}{2} e^{- \mathrm{i} \phi_L / 2} \gamma_L, \\
        a_R & \rightarrow \tfrac{1}{2} e^{- \mathrm{i} \phi_R / 2} \gamma_R,
    \end{split}
\end{align}
thus obtaining the low-energy effective tunnel Hamiltonian
\begin{align}
    \begin{split}
        H_{\text{t,eff}} & = J_L \left( (- \mathrm{i} e^{- \mathrm{i} \phi_L / 2}) a^\dagger_I - (- \mathrm{i} e^{- \mathrm{i} \phi_L / 2})^* a_I \right) \gamma_L \\
        & + J_R \left( (- e^{- \mathrm{i} \phi_R / 2}) a^\dagger_I - (- e^{- \mathrm{i} \phi_R / 2})^* a_I \right) \gamma_R.
    \end{split}
\end{align}
This can be recast in a more convenient form by performing the gauge transformation $a_I \rightarrow (- \mathrm{i} e^{- \mathrm{i} \phi_L / 2}) a_I$, leading to
\begin{align}
    H_{\text{t,eff}} & = J_L (a^\dagger_I - a_I ) \gamma_L + J_R (e^{- \mathrm{i} (\phi_R - \phi_L + \pi) / 2} a^\dagger_I - e^{\mathrm{i} (\phi_R - \phi_L + \pi) / 2} a_I) \gamma_R.
\end{align}
In this form, $H_{\text{t,eff}}$ clearly appears asymmetric with respect to the interchange $\gamma_L \leftrightarrow \gamma_R$ if the phase difference $\Delta \phi \equiv \phi_R - \phi_L$ does not satisfy $\Delta \phi = (4n - 1) \pi$ for some integer $n$. Away from these fine-tuned values of $\Delta \phi$, the tunnel Hamiltonian thus provides a means of distinguishing between the different states of the Majorana qubit composed of $\gamma_L$ and $\gamma_R$, i.e. between different parity sectors of the full system. In order for this information to remain unaccessible to the system---thereby restoring what is usually referred to as the ``topological protection'' or ``parity protection'' of the Majorana qubit---it is therefore necessary to fine-tune the phase difference between the Majorana-like modes to some appropriate value (which can easily be done in our optical setting by adjusting the relative phase of the parametric pumps of each Kitaev chain; see main text). Assuming that this is satisfied---so that the even- and odd-parity sectors of the full system remain degenerate---the low-energy effective tunnel Hamiltonian reduces to
\begin{align}
    H_{\text{t,eff}} & = (a^\dagger_I - a_I) (J_L \gamma_L + J_R \gamma_R) \nonumber \\
    & = J_{LR} (a^\dagger_I - a_I) \gamma_{LR},
\end{align}
where $J_{LR} = \sqrt{J^2_L + J^2_R}$ and $\gamma_{LR} = (J_L \gamma_L + J_R \gamma_R) / J_{LR}$. In this parity-protected scenario, everything therefore happens as if a single Majorana mode $\gamma_{LR}$ was tunnel-coupled to the intermediate cavity. Taking into account the possible detuning $\delta_I$ of the intermediate cavity with respect to the frequency $\omega_p$ of the parametric pumps driving both Kitaev chains (see main text), the low-energy effective Hamiltonian describing the Majorana system tunnel-coupled to the intermediate cavity becomes
\begin{align}
    H_{\text{eff}} & = \delta_I a^\dagger_I a_I + J_{LR} (a^\dagger_I - a_I) \gamma_{LR},
\end{align}
where, of course, $\abs{\delta_I}, J_{LR} \ll E_g$ must be assumed ($E_g$ being the energy gap of the Kitaev chains) in order for the low-energy description to be valid. It was shown in Ref.~\cite{Flensberg11} that the above Hamiltonian allows for a variety of unitary ``tunnel-braid'' operations within the subspace associated with the Majorana modes $\gamma_L$ and $\gamma_R$. In particular, starting from an empty intermediate cavity with $\delta_I > 0$ and $\delta_I / J_{LR} \gg 1$, one can generate an adiabatic transition from an empty to an occupied intermediate cavity and a unitary rotation $U(J_L, J_R) = \gamma_{LR}$ in the Majorana subspace by adiabatically tuning $\delta_I / J_{LR}$ from large positive to large negative values (keeping $\abs{\delta_I}, J_{LR} \ll E_g$). The same unitary operation can be obtained if, starting from an occupied intermediate cavity with $\delta_I < 0$ and $\abs{\delta_I} / J_{LR} \gg 1$, one adiabatically raises $\delta_I / J_{LR}$ from large negative to large positive values. In both cases, the intermediate cavity and the Majorana system become entangled, and an inversion of the occupation of the intermediate cavity as well as a rotation $\gamma_{LR}$ in the Majorana subspace are obtained as the latter return to an unentangled state at the end of the process. Using similar adiabatic processes, one can more generally perform non-Abelian operations within the degenerate ground-state subspace associated with $\gamma_L$ and $\gamma_R$. In particular, one can generate the operation $U_B = \tfrac{1}{\sqrt{2}} (1 + \gamma_L \gamma_R)$ and thus effectively \emph{simulate} the braiding of $\gamma_L$ and $\gamma_R$ in real space (see, e.g., Ref.~\cite{Ivanov01}) following the procedure below:
\begin{enumerate}
    \item Start with $J_L = J_R = 0$ (no tunnel coupling) and the intermediate cavity far blue-detuned ($\delta_I \gg 0$), such that it is initially empty.
    \item Switch on the tunnel coupling $J_L$, raising it to a constant value $J \ll \delta_I$. At this point, the intermediate cavity and the Majorana system are still effectively decoupled.
    \item Adiabatically red-shift the intermediate cavity, taking $\delta / J_{LR}$ from large positive to large negative values. In doing so, the intermediate cavity adiabatically changes from empty to occupied, and the unitary operation $U(J_L = J, J_R = 0) = \gamma_L$ is performed in the Majorana subspace.
    \item Switch on the tunnel coupling $J_R$, raising to the constant value $J_L = J$.
    \item Adiabatically blue-shift the intermediate cavity, taking $\delta / J_{LR}$ from large negative to large positive values. In doing so, the intermediate cavity adiabatically changes from occupied to empty, and the unitary operation $U(J_L = J, J_R = J) = \tfrac{1}{\sqrt{2}} (\gamma_L + \gamma_R)$ is performed in the Majorana subspace.
    \item Switch off both $J_L$ and $J_R$.
\end{enumerate}
At the end of the above cycle, the intermediate cavity and the Majorana system are left unentangled as they were initially. While the intermediate cavity returns to its initial empty state, a non-trivial unitary operation $U_B = \tfrac{1}{\sqrt{2}} (\gamma_L + \gamma_R) \gamma_L = \tfrac{1}{\sqrt{2}} (1 + \gamma_L \gamma_R)$ is obtained in the Majorana subspace, as if $\gamma_L$ and $\gamma_R$ had physically been exchanged. This exemplifies the fact that one can \emph{simulate} real-space braiding in purely one-dimensional systems. We remark, however, that the tunnel-braid operations used in the above procedure are only parity-protected provided that the phase difference between the Majorana modes $\gamma_L$ and $\gamma_R$ satisfies $\Delta \phi = (4n - 1) \pi$ (for some integer $n$). Because of this fine-tuning requirement, tunnel-braid operations are not strictly speaking topologically protected. Moreover, even if $\Delta \phi$ is fine-tuned so that the degeneracy of the Majorana modes is preserved, one must still be able to control the tunneling amplitudes $J_L$ and $J_R$ sufficiently well in order to ensure that the desired tunnel-braid operations are generated. Indeed, different relative strengths $J_L / J_R$ generally lead to different operations (see Ref.~\cite{Flensberg11} for further details). Although the lack of topological protection renders tunnel-braid operations somewhat unattractive for solid-state proposals where fermion parity conservation is almost perfect, it does not constitute an additional problem in our optical setting where parity is anyway only conserved on timescales much shorter than the lifetime of the photons. If tunnel-braid operations can be completed within such timescales, the non-Abelian nature of Majorana-like modes of light can be simulated.

